\documentclass{nature}
\usepackage{amsmath,amsfonts,amssymb,times,graphicx,color,lineno,caption}
\usepackage[nolists,nomarkers]{endfloat}

\title{An experimental test of the viscous anisotropy hypothesis for
  partially molten rocks}
\author{Chao Qi$^{1}$, David L.~Kohlstedt$^1$ Richard F.~Katz$^2$ \&
  Yasuko Takei$^3$}

\begin{document}

\maketitle

\begin{affiliations}
 \item Department of Earth Sciences, University of Minnesota, Minneapolis, Minnesota 55455, USA.
 \item Department of Earth Sciences, University of Oxford, Oxford OX1 3AN, UK.
 \item Earthquake Research Institute, University of Tokyo, Tokyo 113-0032, Japan
\end{affiliations}

%\linenumbers

\begin{abstract}
  Chemical differentiation of rocky planets occurs by melt segregation
  away from the region of melting.  The mechanics of this process,
  however, are complex and incompletely understood.  In partially
  molten rocks undergoing shear deformation, melt pockets between
  grains align coherently in the stress
  field\cite{kohlstedt1996b,zimmerman1999}; it has been hypothesized
  that this anisotropy in microstructure creates an anisotropy in the
  viscosity of the aggregate\cite{takei2009a}.  With the inclusion of
  anisotropic viscosity, continuum, two-phase-flow models reproduce
  the emergence and angle of melt-enriched bands that form in
  laboratory experiments\cite{takei2009c, takei2013}.  In the same
  theoretical context, these models also predict sample-scale melt
  migration due to a gradient in shear stress.  Under torsional
  deformation, melt is expected to segregate radially
  inward\cite{takei2013, katz2013}.  Here we present new torsional
  deformation experiments on partially molten rocks that test this
  prediction.  Microstructural analyses of the distribution of melt
  and solid reveal a radial gradient in melt fraction, with more melt
  toward the centre of the cylinder.  The extent of this radial melt
  segregation grows with
  progressive strain, consistent with theory.  The agreement between
  theoretical prediction and experimental observation provides a
  validation of this theory, which is critical to understanding the
  large-scale geodynamic and geochemical evolution of Earth.
\end{abstract}

Shear deformation of partially molten rocks gives rise to melt
segregation into sheets (bands in cross-section) that emerge at a low
angle to the shear plane.  This mode of segregation was predicted with
two-phase flow theory\cite{stevenson1989} and subsequently discovered
in experiments\cite{holtzman2003a, king2010}.  It has been proposed
that melt-enriched bands, if present in the mantle of Earth, would
permit rapid extraction of melt\cite{kohlstedt2009}, produce
significant anisotropy in seismic wave propagation\cite{kendall1994},
and provide a mechanism for the seismic discontinuity that is, in some
places, associated with the lithosphere--asthenosphere
boundary\cite{kawakatsu2009}.  The emergence\cite{spiegelman2003} and
low angle\cite{katz2006} of melt-enriched bands under simple-shear
deformation can be reproduced using two-phase flow theory with a
non-Newtonian, isotropic viscosity.  This theory describes the flow of
a low-viscosity liquid (melt) through a permeable and viscously
deformable solid matrix (grains)\cite{mckenzie1984}.  However, an
unrealistically strong stress-dependence of viscosity was required to
match the low angle of bands observed in experiments\cite{katz2006}.
This disagreement between models and experiments found a possible
resolution by the incorporation of anisotropic viscosity arising from
coherent alignment of melt pockets between grains (i.e.,
melt-preferred orientation, MPO) in response to a deviatoric
stress\cite{takei2009a, takei2009b,butler2012, takei2013}.

Crucially, with the inclusion of viscous anisotropy, two-phase flow
theory also predicts a simultaneous but distinct mode of melt
segregation driven by large-scale gradients in shear stress.  This
mode is termed base-state melt segregation\cite{takei2009c, takei2013,
  katz2013}.  Base-state melt segregation is not predicted if
viscosity is isotropic; thus, its occurrence in experiments represents
a test of the hypothesis that MPO leads to anisotropy in viscosity.
Below we explain base-state melt segregation in more detail; we then
present new experimental results that demonstrate its occurrence.

\begin{figure}
    \centering
    \includegraphics[width=8.9cm]{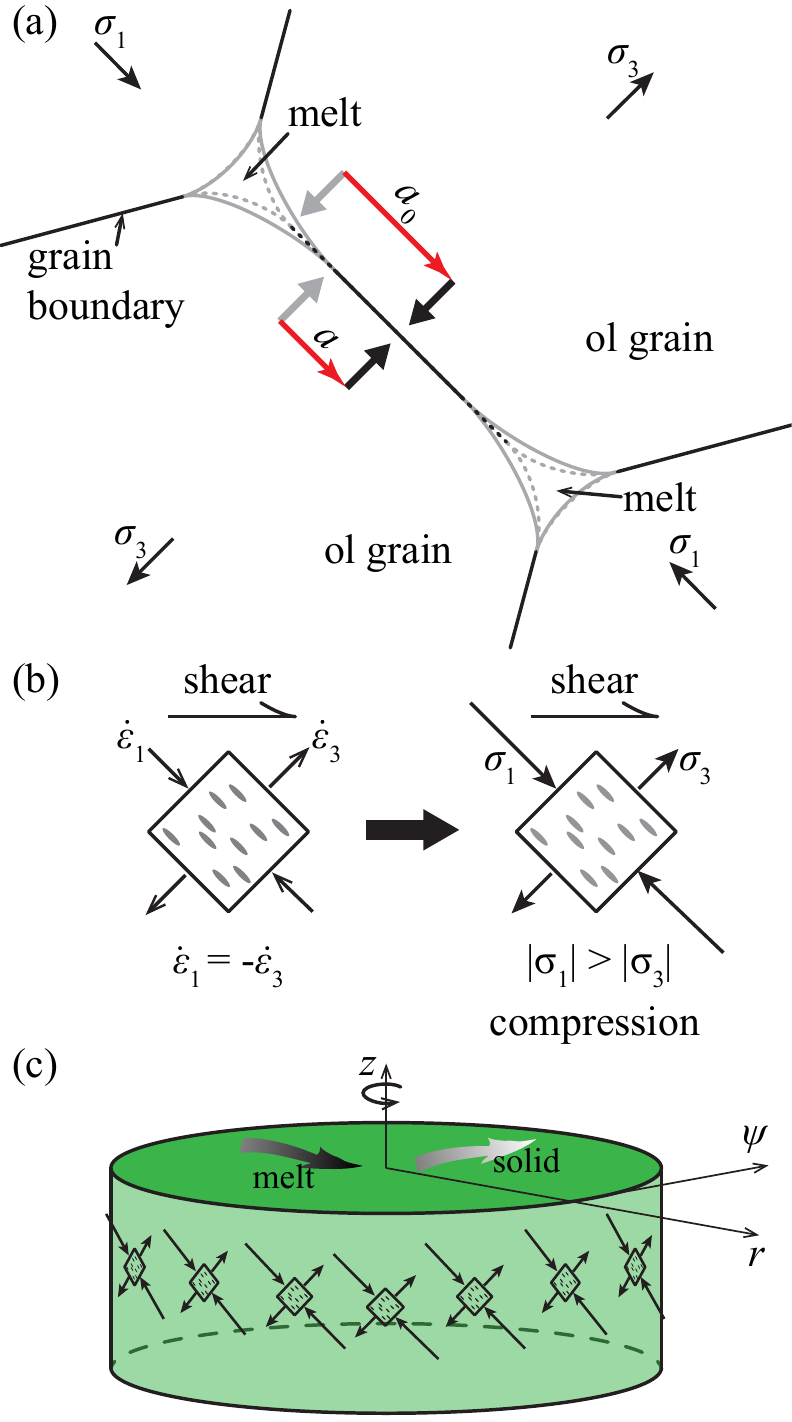}
    \caption{\textbf{Schematic diagrams of the interactions between
        stress and anisotropic viscosity.} \textbf{a}, Stress-induced
      melt redistribution and its influence on diffusion pathways in
      2D. The grain boundaries in contact with a neighbouring grain
      are in black, while the grain boundaries in contact with melt
      are in gray. Melt pockets under isotropic conditions are
      outlined with dash lines, while the redistributed melt pockets
      under shear deformation are outlined with solid lines. Red
      arrows show the lengths of diffusion pathways from the
      grain-melt boundary (gray arrows) to the centre of a grain
      boundary (black arrows). The diffusion pathway shortens from
      $a_0$ to $a$ with applied deviatoric stress
      $\sigma_3$. \textbf{b}, The coupling between torsional
      deformation with zero volumetric strain rate and stress with
      non-zero volumetric component associated with viscous
      anisotropy. The square is a REV with melt pockets (gray) aligned
      45$^\circ$ to the shear plane, antithetic to the shear
      direction. Modified from previous
      study\cite{takei2013}. \textbf{c}, A sample-scale view of the
      development of the hoop stress from compressive stresses due to
      cylindrical geometry.}
    \label{fig:theory}
\end{figure}

Figure~\ref{fig:theory}a illustrates how anisotropy in grain/melt
microstructure (i.e., MPO) produces viscous anisotropy through the
mechanics of diffusion creep\cite{takei2009a}.  For a representative
grain in an aggregate subjected to a deviatoric stress, the contact
area with neighbouring grains decreases for grain boundaries that are
normal to the direction of the minimum principal stress $\sigma_3$
(the minimum eigenvalue of the deviatoric stress tensor; compression
positive).  A decrease in contact area shortens the diffusion pathway
for material transport along this boundary relative to grain
boundaries with different orientations.  Since melt provides a fast
pathway for diffusion, this change in the contiguity between
neighbouring grains reduces the timescale of the diffusive response to
the $\sigma_3$ component of stress.  Conversely, an increase in
grain--grain contact area in the direction of the maximum principal
stress $\sigma_1$ lengthens the diffusion pathway on this surface and
thus the timescale for diffusive response to stress in this direction.

To translate these concepts from the microscopic to the continuum
scale, consider a representative element of volume (REV) that is large
enough to contain many microscopic units (grains and melt pockets) and
small enough to define a point property at the scale of macroscopic
features of interest (Fig.~\ref{fig:theory}b).  Under deviatoric
stress, melt pockets in the REV coherently align normal to the
$\sigma_3$ direction and the timescale for the diffusive response to
stress is reduced in this direction.  If the dominant deformation
mechanism of the aggregate is diffusion creep, this rapid response
imparts a reduction of the continuum viscosity in the
$\sigma_3$ direction\cite{cooper1986, takei2009a}.  Likewise, the
change in grain contiguity associated with $\sigma_1$ increases the
viscosity in that direction.

Viscous anisotropy can be quantified with a highly symmetric,
fourth-rank tensor (Supp.~Mat.).  The orientation of this tensor is
described by three angles that rotate it with respect to the system
coordinates. This rotation is used to align anisotropy with the
principal directions of deviatoric stress.  As a simplifying
approximation, we assume that, at each point in the domain, the plane
containing $\sigma_1$ and $\sigma_3$ is parallel to the imposed shear
direction and perpendicular to the imposed shear plane.  This leaves
only one angle to be determined, the angle $\Theta$ between the shear
plane and the $\sigma_3$ direction.  The magnitude of anisotropy is
parameterised with two scalars: $\alpha$ specifies the viscosity
reduction in the $\sigma_3$ direction; $\beta$ specifies the viscosity
increase in the $\sigma_1$ direction.  When either or both $\alpha$
and $\beta$ are non-zero at a point in the continuum, the viscosity at
that point is anisotropic.  The associated tensor then has non-zero
off-diagonal terms that couple shear stress to normal strain rate (and
vice versa).  It is these terms that give rise to base-state
segregation\cite{takei2009c, takei2013}.

To clarify the physical mechanism of base-state segregation, consider
a cylindrical sample in a sealed chamber, deformed in torsion at a
constant twist rate (Fig.~\ref{fig:theory}c).  Before any deviatoric
stress is applied, the grain/melt microstructure is isotropic and the
rate of (de)compaction is zero everywhere within the sample.  With
initiation of twisting, as a consequence of the deviatoric stress, a
MPO develops and the viscosity becomes anisotropic.  The imposed
strain rate, aligned melt pockets, and consequent pattern of stress
are shown schematically in Figure~\ref{fig:theory}b--c.  The disparity
between $|\sigma_1|$ and $|\sigma_3|$ gives rise to a net compression
that, because it is everywhere tangent to the cylinder, is a
compressive hoop stress.  This compressive hoop stress pushes the
solid grains radially outward and causes a pressure gradient that
drives melt radially inward\cite{takei2013} (details provided in
Supp.~Mat.).  This differential motion is the base-state melt
segregation under torsional deformation.

\begin{figure}
    \centering
    \includegraphics[width=7cm]{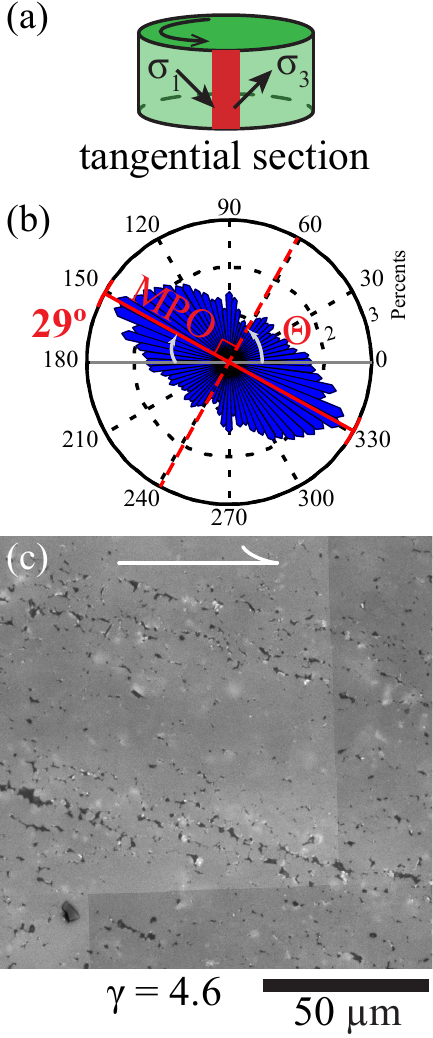}
    \caption{\textbf{MPO in a sample deformed to a shear strain of
        4.6.}  \textbf{a}, A sketch illustrating a deformed
      cylindrical sample with its tangential section marked in red.
      Arrows in this sketch illustrate the expected directions of
      deviatoric stresses $\sigma_1$ and $\sigma_3$ in tangential
      section.  \textbf{b}, A rose diagram generated from optical
      micrographs of the tangential section at a local shear strain of
      $\gamma= 4.6$ (PI0817).  The rose diagram was constructed from
      more than 1000 melt pockets with areas larger than 2 $\mu$m$^2$.
      The length of each petal represents the ratio of melt pockets
      with this certain orientation, scaled by percentage.  $\Theta$
      is anisotropy angle, which the angle between red, dashed lines
      and 0$^{\circ}$.  \textbf{c}, Optical micrograph from the
      tangential section of the sample.  Olivine grains are light
      gray; melt is dark gray; internal reflections or residual
      polishing material are white.  The sense of shear is marked by
      the arrow on top.  }
    \label{fig:rose}
\end{figure}

\begin{figure}
    \centering
    \includegraphics[width=15cm]{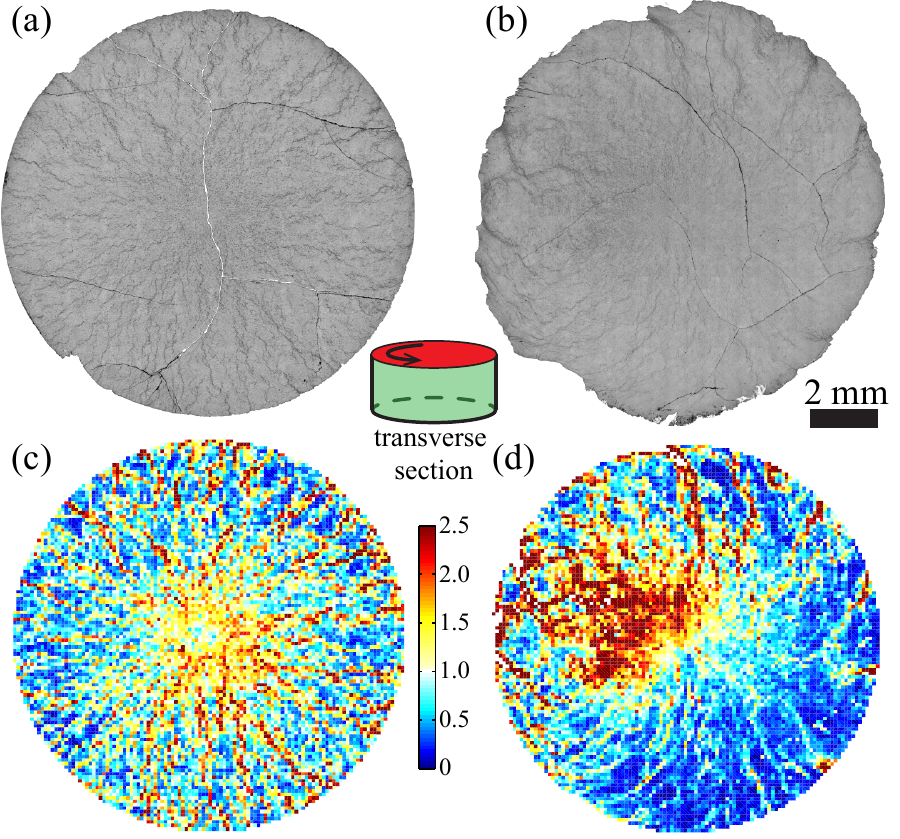}
    \caption{\textbf{Optical micrographs and processed images
        demonstrating base-state melt segregation.}  \textbf{a} and
      \textbf{b}, Optical micrographs of transverse sections from
      samples sheared to $\gamma(R) =$ 5.0 (PI0817) and 14.3 (PI0891),
      respectively.  Olivine is light gray and melt is dark gray.  The
      sketch located between \textbf{a} and \textbf{b} is a deformed
      cylindrical sample with its transverse section marked in red.
      \textbf{c} and \textbf{d}, Melt distribution maps generated from
      \textbf{a} and \textbf{b}, respectively.  The grid size is 100
      $\times$ 100 $\mu$m.  Color bar indicates melt fraction
      normalized to the average melt fraction in the image.  Due to
      its high strain, sample PI0891 sheared off-axis so that the
      torsional axis is to the northwest of the centre of the image. }
    \label{fig:maps}
\end{figure}

To test this prediction and hence the hypothesis that viscosity is
anisotropic, we imposed a constant twist-rate on cylindrical samples
of partially molten rock that initially had uniform melt fraction
(Table~\ref{tab:experiments}).  In tangential sections of quenched
samples that were deformed in torsion (Fig.~\ref{fig:rose}), we
observe aligned melt pockets and low angle, melt-enriched bands.
Melt-enriched bands are also evident in transverse sections
(Fig.~\ref{fig:maps}).  More importantly, analyses of optical
micrographs of transverse sections reveal a gradient in melt fraction
in the radial direction, with melt concentrated toward the axis of the
cylinder.  This gradient in melt fraction corresponds to the
base-state melt segregation predicted if viscosity is anisotropic.
Our observations of MPO, melt-enriched bands, and radial melt
segregation are detailed in subsequent paragraphs.

The rose diagram in Figure~\ref{fig:rose}b demonstrates that at a
local shear strain of $\gamma= 4.6$, melt pockets are aligned at
$\sim$29$^{\circ}$ to the shear plane, antithetic to shear direction.
In contrast, the expected $\sigma_3$ direction, based on cylindrical
simple shear flow with isotropic viscosity, is 45$^\circ$ to the shear
plane.  The observed low angle of melt alignment means that, at this
shear strain, either melt pockets are not normal to the $\sigma_3$
direction\cite{zimmerman1999} or $\sigma_3$ has rotated
counter-clockwise.  The reason for this alignment is unknown; it might
be due to the emergence of chains of melt pockets\cite{holtzman2007}
(Fig.~\ref{fig:rose}c) or to the anisotropic viscosity itself.
In the theory of two-phase flow with viscous anisotropy, elaborated in
the Supplementary Material, it is generally assumed that melt pockets
align perpendicular to $\sigma_3$, as suggested by deformation
experiments on an analogue material at small strains ($\gamma <
0.2$)\cite{takei2010}.  The observed MPO, therefore, may represent a
subtle but important discrepancy between observation and theory that
we return to below.  Despite this possible discrepancy, the observed,
strong MPO demonstrates the microstructural anisotropy that
hypothetically causes viscous anisotropy.

Two-phase flow theory with anisotropic viscosity\cite{takei2013} also
predicts the emergence of sheets of high melt fraction that appear as
bands in two-dimensional sections\cite{king2010}.  In
Figure~\ref{fig:maps}, these features appear as radial lines of high
melt fraction where sheets cross the transverse section.  For the
sample deformed to an outer-radius shear strain of $\gamma(R)= 5.0$
(Fig.~\ref{fig:maps}a,~c), the melt-enriched bands are distributed
uniformly around the cylinder, whereas at a larger strain of $\gamma
(R)= 14.3$ (Fig.~\ref{fig:maps}b,~d), the azimuthal distribution of
melt-enriched bands is inhomogeneous, dominated by several
extraordinarily large bands.  Because the total strain decreases
toward the centre of the cylinder, the region close to axial centre
exhibits less banding.  More significantly, however,
Figures~\ref{fig:maps}c and d demonstrate a general increase in melt
fraction toward the centre of the cylinder, consistent with the
predicted base-state migration of melt radially inward.

\begin{figure}
    \centering
    \includegraphics[width=12cm]{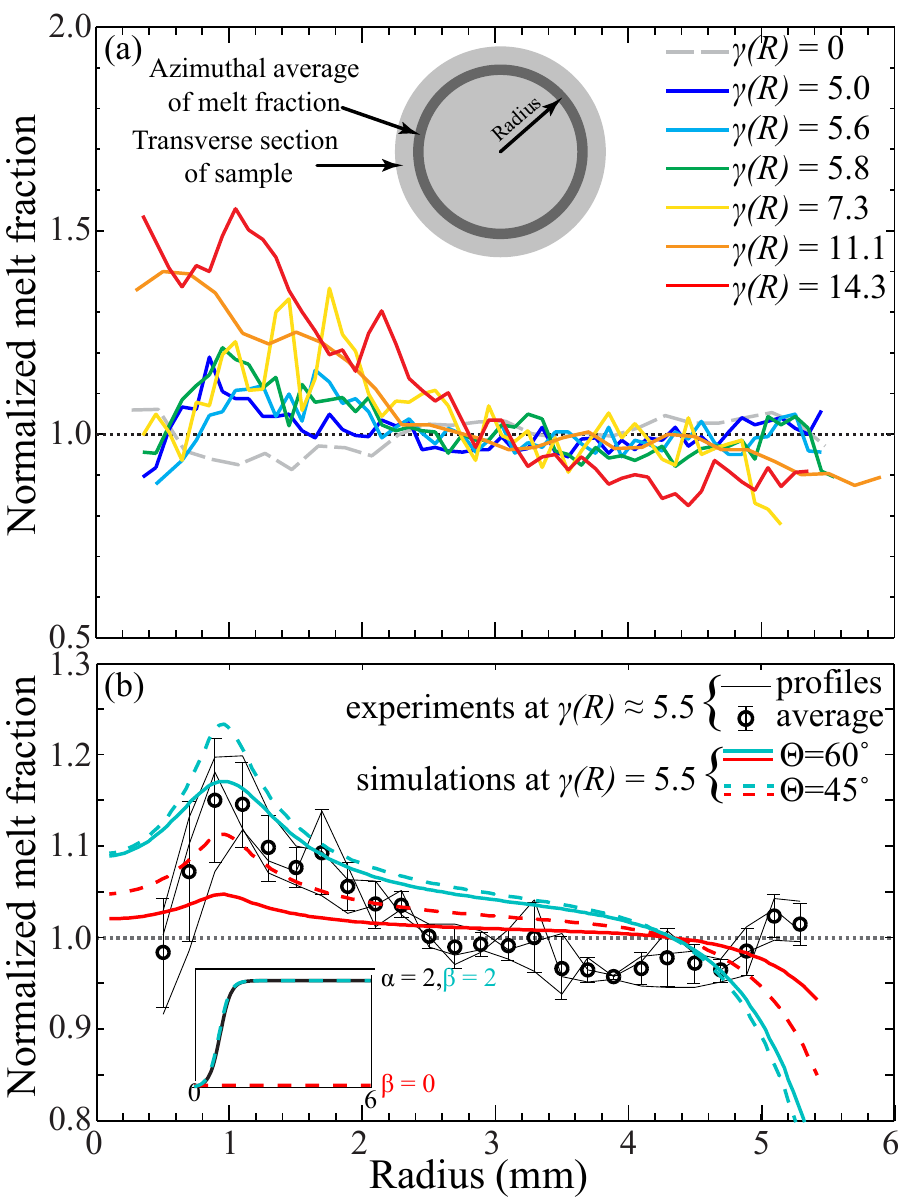}
    \caption{\textbf{Azimuthally averaged, normalized melt fraction
        versus radius.} \textbf{a}, Plot of azimuthally averaged,
      normalized melt fraction versus radius for starting material and
      six deformed samples. In the legend, the outer-radius strain of
      each sample is noted. Melt fraction is normalized by the average
      value of the transverse section of each sample. The black dotted
      line corresponds to a normalized melt fraction of 1.0. The
      sketch shows how an azimuthal average of melt fraction is
      obtained for a given radius. \textbf{b}, Reproducibility test
      for three samples sheared to $\gamma(R) = 5.5\pm0.5$ and its
      comparison with the results from numerical simulations.
      Simulations were conducted with an initial compaction length of
      $\delta_c=0.1R$, a bulk-to-shear viscosity ratio of $r_\xi=10$,
      anisotropy angles of $\Theta=$ 60$^\circ$ and 45$^\circ$, and
      anisotropy magnitudes $\alpha(r)$ and $\beta(r)$ shown in the
      inset panel.  Blue lines are for the case with
      $\alpha_\textrm{max}= 2$ and $\beta_\textrm{max}=2$, while red
      lines are for the case with $\alpha_\textrm{max}=2$ and
      $\beta=0$.  There was no initial porosity perturbation and hence
      no band formation.  Further details are provided in the online
      supplementary materials.}
    %% OTHER PARAMETERS: $\phi_0=0.1$, $\lambda=27$, $n=3$,
    \label{fig:profile}
\end{figure}

Radial profiles of the azimuthally averaged, normalized melt fraction
are presented in Figure~\ref{fig:profile}a for seven experiments, each
with a different final strain.  The melt fraction in an experiment
with no deformation (grey line) varies by less than 10\% along a
radius.  In all deformed samples, the melt fraction increases toward
the centre of the cylinder --- evidence for base-state segregation.
For the three samples deformed to an outer-radius shear strain of
$\gamma(R) = 5.5 \pm 0.5$, each radial profile of melt concentration
reaches its peak at a radius of $r^\textrm{peak} \approx 1$~mm,
corresponding to a shear strain of $\gamma(r^\textrm{peak}) \approx
1$.  Melt fraction decreases from that point toward the axis of the
cylinder; this behavior is expected because the low-stress/low-strain
region at small radius has little or no MPO and hence has essentially
isotropic viscosity ($\alpha=\beta= 0$).  For samples deformed to
higher outer-radius shear strains ($\gamma(R) = 7.3,\,11.1,$ and
$14.3$), peaks in melt fraction occur at a radius of $r^\textrm{peak}
< 0.2$~mm.  The sample with the highest outer-radius shear strain
($\gamma(R) =14.3$) exhibits the largest ratio of maximum to minimum
melt fraction $\phi_\textrm{max}/\phi_\textrm{min}$
(Table~\ref{tab:experiments}), a measure of the strength of base-state
melt segregation.  Except for the samples sheared to outer-radius
shear strains of $\gamma(R) = 5.0$ and $7.3$, the maximum in melt
fraction increases with increasing shear strain.  In summary, the
results presented in Figure~\ref{fig:profile}a demonstrate that, with
increasing strain, the pressure gradient induced by anisotropic
viscosity drives melt inward, increasing the maximum value of the
azimuthally averaged melt fraction and decreasing the radius at which
this maximum occurs.

Figure~\ref{fig:profile}b compares the azimuthally averaged profiles
of normalized melt fraction from three samples deformed to $\gamma(R)
= 5.5\pm0.5$ with those derived from numerical simulations.  The data
points in \ref{fig:profile}b, which are the mean values of the of the
azimuthal averages at each radius, reach a maximum normalized melt
fraction of $\sim$1.15 at $r^\textrm{peak}\approx 1$~mm and a minimum
of $\sim$0.95 at $r \approx 4$~mm.  For comparison, radial profiles of
melt fraction from numerical simulations of samples deformed to
$\gamma(R)=5.5$ at an initial compaction length of $\delta_c=0.1R$ and
a bulk-to-shear viscosity ratio of $r_\xi=10$.  In the simulations,
two conditions are used for the angle of viscous anisotropy: (1)
$\Theta = $ 45$^{\circ}$, suggested by previous
experiments\cite{takei2010}, and (2) $\Theta =$ 60$^{\circ}$,
suggested by Figure \ref{fig:rose}.  The other variable in the
simulations is the magnitude of viscous anisotropy.  In all four
simulations, $\alpha$ increases from zero at the centre of the
cylinder to $\alpha_\textrm{max} =2$ at $r\approx 1$~mm and then
remains constant at larger radii.  In two of the simulations, $\beta$
mimics the behavior of $\alpha$.  In the other two simulations,
$\beta$ is zero at all radii.  In the decompaction region (i.e. at
small radii), profiles with $\Theta =$ 45$^{\circ}$ exhibit higher
melt fractions than those with $\Theta =$ 60$^{\circ}$, while profiles
with $\alpha_\textrm{max} = \beta_\textrm{max} = 2$ exhibit higher
melt fractions than those with $\alpha_\textrm{max} = 2$ and
$\beta=0$.  The profile with $\Theta =$ 60$^{\circ}$ and
$\alpha_\textrm{max} = \beta_\textrm{max} = 2$ is in the most
consistent with the experimental results.  However, some clear
differences exist between the simulated and the experimental profiles.
First, for $2.6<r<4.4$ mm the simulated profiles lie above
experimentally measured profile.  Second, the abrupt decrease in the
simulated profiles at $r>4.4$ mm was not observed experimentally.
Despite these quantitative discrepancies, all simulated porosity
profiles are in good qualitative agreement with those from
experiments, in terms of both the amplitude and the radial position of
the porosity maximum.

In this paper we presented experimental observations of the radial
distribution of melt in partially molten rocks deformed in torsion to
large strain.  For this deformation geometry, the theory of melt
segregation with anisotropic viscosity predict a radial distribution
of melt fraction. The inclusion of viscous anisotropy in the theory is
a necessary and sufficient condition for the development of radially
inward, base-state melt segregation.  Our experiments test this
prediction, and the results reported here are in general agreement
with theory, validating the viscous-anisotropy hypothesis.  This
experimental validation of MPO-induced viscous anisotropy represents a
significant advance in our understanding of the relationship between
microstructure and continuum mechanics of partially molten rocks, and
it also exposes details of the linkage between deformation, MPO and
viscosity that are not captured by present models.  In Earth,
partially melting largely occurs in regions of active deformation ---
places where melt-preferred orientation will occur.  Inclusion of
viscous anisotropy in models of the dynamics of the partial molten
mantle will likely predict shear-induced melt migration that
profoundly influences melt segregation, mantle dynamics, and chemical
evolution of our planet.

\begin{methods}
  Samples were fabricated from mixtures of fine-grained powders of
  olivine from San Carlos, AZ, plus 10 vol.\% alkali basalt from
  Hawaii\cite{morgan2003}.  Olivine powders were obtained by grinding
  San Carlos olivine crystals in a fluid-energy mill to produce a
  particle size of 2 $\mu$m.  Before mechanically mixing with alkali
  basalt powders with a particle size of $\sim$10 $\mu$m, the olivine
  powders were dried at 1373 K for 12 h at an oxygen partial pressure
  near the Ni-NiO buffer to remove water and carbon-based impurities
  introduced during the grinding process.  Mixtures were uniaxially
  cold-pressed at 100 MPa into nickel capsules and then
  hydrostatically hot-pressed at 1473 K and 300 MPa for 3.5 h in a
  gas-medium apparatus\cite{paterson2000}.  After hot-pressing,
  samples were cut into thin cylinders with a diameter of $2R \approx
  12$ mm and a thickness of 3 to 5 mm.  The cut sample was then placed
  into a nickel capsule with spacers cored from a coarse-grained
  natural dunite as end caps, thus providing non-reactive, impermeable
  boundaries during deformation\cite{qi2013}.  The sample, Al$_2$O$_3$
  spacers and pistons, and ZrO$_2$ pistons were enclosed in an iron
  jacket for deformation.

  Torsion experiments were conducted at a shear strain rate of
  10$^{-3.5}$ s$^{-1}$, a temperature of 1473 K, and a confining
  pressure of 300 MPa in a gas-medium apparatus fitted with a torsion
  actuator\cite{paterson2000}.  After achieving the target strain,
  each sample was cooled rapidly ($\sim$2 K/s) to 1300 K under the
  torque imposed at the end of the deformation experiment to preserve
  the deformation-produced microstructure and then cooled to room
  temperature with no torque applied.  After deformation, with the
  iron jacket and the nickel capsule dissolved in acid, the deformed
  sample was cut in half perpendicular to torsional axis, leaving two
  transverse sections for examinations.  Each transverse section was
  polished on a series of diamond lapping films down to 0.5 $\mu$m,
  followed by a final step using colloidal silica.  The section was
  then examined by reflected-light optical microscopy after chemically
  etching with diluted HF to highlight melt pockets.

  To map the whole transverse section with an area of $\sim$113
  mm$^2$, a mosaic image consisting of 2209 high-resolution (0.3
  $\mu$m per pixel) optical micrographs was used.  A binary image with
  melt appearing white was created from this mosaic image using using
  a combined image segmentation method, which includes edge
  detection\cite{canny1986,lim1990,parker2010} and a threshold of
  grayscale.  Then a profile of melt fraction was calculated from the
  area fraction of the white pixels.
\end{methods}

% use the bibtex bibliography database manuscript.bib
\bibliographystyle{style/naturemag}
\bibliography{manuscript}

\begin{addendum}
\item The authors are thankful to Mark Zimmerman and Mat\v{e}j Pe\v{c}
  for helps with experiments and analyses, and to Yan Liang and Clint
  Conrad for providing the alkali basalt.  This study is supported by
  NSF grant EAR-1214876.  Katz is grateful to the Leverhulme Trust for
  support.  Numerical models were run on ARCHER, the UK national
  supercomputer.
 \item[Competing Interests] The authors declare that they have no
competing financial interests.
 \item[Correspondence] Correspondence and requests for materials
should be addressed to C.~Qi (email: qixxx063@umn.edu).
\end{addendum}

\begin{table}[ht]
  \centering
  \begin{tabular}{c|cccc}
    Sample & $\gamma(R)$ & $\dot{\varepsilon}_{eq}$ (s$^{-1}$) & $\sigma_{eq}$ (MPa) & $\phi_{max} / \phi_{min}$ \\\hline
    PI0767 & 11.1 & 2.29$\times$10$^{-4}$ & 187 & 1.6\\
    PI0811 & 5.6 & 1.84$\times$10$^{-4}$ & 237 & 1.2\\
    PI0812 & 5.8 & 1.84$\times$10$^{-4}$ & 163 & 1.4\\
    PI0817 & 5.0 & 2.35$\times$10$^{-4}$ & 197 & 1.3\\
    PI0839 & 7.3 & 1.84$\times$10$^{-4}$ & 237 & 1.7\\
    PI0891 & 14.3 & 2.04$\times$10$^{-4}$ & 179 & 2.0
  \end{tabular}
  \caption{\textbf{Experiments Summary.} $\gamma(R)$ is the 
    outer-radius shear strain, where $R$ is the radius of a 
    sample with   $R \approx$ 6 mm.
    The equivalent strain rate and stress, $\dot{\varepsilon}_{eq}$ 
    and $\sigma_{eq}$, respectively, are calculated from shear strain
    rate and stress using Cauchy stress tensor.
    $\phi_{max} / \phi_{min}$ is the ratio of maximum  to minimum melt 
    fraction in the profile of azimuthally averaged melt fraction.}
  \label{tab:experiments}
\end{table}
\clearpage

\section*{Online Supplementary Material}
\paragraph{The viscosity tensor} The diffusion-creep viscosity tensor
$C_{ijkl}$ of a partially molten rock can be calculated by using a
microstructure-based model\cite{takei2009a}.  The grain coordinates
$(x_g, y_g, z_g)$ are defined independently of the continuum
coordinates $(x, y, z)$.  Each grain in the aggregate is assumed to
have 14 circular contacts with equal radius.  Microstructural
anisotropy is represented by a decrease (or increase) in the radius of
the two contact faces in the $x_g$ (or $y_g$) direction. By
substituting this contact geometry into equation (42) of Takei \&
Holtzman\cite{takei2009a}, $C_{ijkl}$ $(i,j,k,l=x_g,y_g,z_g)$ is
calculated as
\begin{equation}
  C_{ijkl}
  =\xi\,\delta_{ij}\delta_{kl} +
  \eta\left(\delta_{ik}\delta_{jl}+\delta_{il}\delta_{jk}
    -\frac{2}{3}\delta_{ij}\delta_{kl}\right)
  -\Delta \,\delta_{ix_g}\delta_{jx_g}\delta_{kx_g}\delta_{lx_g}
   +\Delta^{\prime} \,\delta_{iy_g}\delta_{jy_g}\delta_{ky_g}\delta_{ly_g},
  \label{eq:Cgc}
\end{equation}
where $\eta$ and $\xi$ correspond to the shear and bulk viscosity of
an isotropic matrix, and $\Delta$ (or $\Delta^{\prime}$) represents the
decrease (or increase) in viscosity in the $x_g$ (or $y_g$) direction.

For illustration, we consider a two dimensional problem in which
contiguity is increased in the $\sigma_1$ direction and decreased in
the $\sigma_3$ direction (in other words, MPO is aligned according to
the instantaneous directions of principal stress). Furthermore, we
assume that the $\sigma_1$--$\sigma_3$ plane is parallel to the
$x$--$y$ plane.  Let $\Theta$ be the angle between the $x_g$ and $x$
axes.  Then, by a coordinate transformation\cite{takei2013} of
expression (\ref{eq:Cgc}), the viscosity tensor $C_{ijkl}$ in the
continuum coordinates $(i,j,k,l=x,y,z)$ is
\begin{eqnarray}
  &&C_{ijkl}  = \eta_0\textrm{e}^{-\lambda(\phi-\phi_0)} \times\nonumber\\
  && \bordermatrix{
    ij\!\downarrow &kl\!\rightarrow xx&yy&zz&yz&zx&xy  \cr
    xx&r_\xi+\frac{4 }{3} -\frac{\alpha+\beta}{2}  \cos2\Theta & r_\xi-\frac{2}{3} &r_\xi-\frac{2}{3}&0
    &0&-\frac{\alpha+\beta}{4}  \sin2\Theta         \cr
    &-\frac{\alpha-\beta}{8} (3+ \cos4\Theta)
    &-\frac{\alpha-\beta}{8} (1- \cos4\Theta) & & & &-\frac{\alpha-\beta}{8}  \sin4\Theta   \cr
    yy& \cdot  & r_\xi+\frac{4 }{3}+\frac{\alpha+\beta}{2}  \cos2\Theta    &  r_\xi-\frac{2}{3}
    &0   & 0&-\frac{\alpha+\beta}{4}  \sin2\Theta       \cr
    &    &-\frac{\alpha-\beta}{8} (3+ \cos4\Theta)& && &+\frac{\alpha-\beta}{8}  \sin4\Theta   \cr
    zz&\cdot  & \cdot  &  r_\xi+\frac{4}{3} &0
    & 0&0     \cr
    yz&\cdot & \cdot&\cdot & 1   &   0&0    \cr
    zx&\cdot & \cdot&\cdot  & \cdot   &   1&0 \cr
    xy&\cdot &\cdot&\cdot  & \cdot   &   \cdot&1-\frac{\alpha-\beta}{8}    (1- \cos4\Theta)  \cr}
  \nonumber\\
  \label{eq:Csystem}
\end{eqnarray}
where only 21 of the 81 components are shown; the others are readily
obtained from the symmetry of $C_{ijkl}$.  The factor in front of the
matrix represents the shear viscosity $\eta$ that decreases
exponentially with increasing melt fraction $\phi$. $\eta_0$
represents $\eta$ at reference melt fraction $\phi_0$, and $\lambda$
is the porosity weakening factor.  The parameter $r_\xi=\xi / \eta$
represents the bulk-to-shear viscosity ratio.  Parameters $\alpha
=\Delta/\eta$ and $\beta=\Delta^\prime/\eta$ represent the magnitude
of viscous anisotropy in the reduced and increased directions,
respectively.  Parameters $r_\xi$, $\alpha$, and $\beta$ are assumed
to be independent of melt fraction based on the theoretical
result\cite{takei2009a, takei2013}.

When $\Delta^\prime=0$ and $\beta=0$, (\ref{eq:Cgc}) and
(\ref{eq:Csystem}) are equal to those used in previous
studies\cite{takei2013, katz2013}.  The present formulation is more
general than that one because it allows for the effect of contiguity
change in the $y_g$ direction as well as the $x_g$ direction.  In
obtaining solutions to the governing equations, $\Theta$ is usually
taken such that the $x_g$ axis is in the $\sigma_3$ direction, as
discussed above.  However, because $\Theta$ is defined by the
direction of microstructural anisotropy independently of the resultant
stress, it can be used to describe a general MPO direction
(Fig.~\ref{fig:rose}).

\paragraph{Base-state segregation in torsion} The anomalous character
of the base-state melt segregation in torsion bears some elaboration.
Takei \& Katz\cite{takei2013} showed that viscous anisotropy has a
general tendency to cause melt segregation up a gradient in shear
stress, and demonstrated the occurrence of base-state melt segregation
directly driven by a gradient of shear stress.  In torsion, shear
stress increases with increasing radius.  However, the inward melt
flow is driven by the negative hoop stress, which is opposite in
direction to general tendency.  A detailed explanation for this
anomalous result for torsion deformation is presented in Takei \& Katz
(\cite{takei2013}, Section 5.2).  Irrespective of these subtleties,
viscous anisotropy plays an essential role in the generation of the
negative hoop stress that drives base-state segregation (Fig.~1).  The
occurrence of radially inward melt segregation in torsion can
therefore be appropriately used to test the viscous anisotropy
hypothesis.

\end{document}